\documentstyle[eqsecnum,12pt,aps]{revtex}
\newcommand{\disc}{\displaystyle}

\newcommand{\lhisb}{{\cl}_{{\disc\chi}\mbox{sb}}}

\newcommand{\vecr}{\vec{r}}

\newcommand{\fpi}{F_{\pi}}

\newcommand{\cl}{\cal {L}}
\newcommand{\bra}[1]{\langle{#1}\vert}
\newcommand{\ket}[1]{\vert{#1}\rangle}

\newcommand{\gzero}{g^{\prime}_0}

\newcommand{\veps}{\varepsilon}

\newcommand{\be}{\begin{equation}}
\newcommand{\ee}{\end{equation}}
\newcommand{\ba}{\begin{array}{l}}
\newcommand{\banonum}{\begin{eqnarray*}}
\newcommand{\ea}{\end{array}}
\newcommand{\eanonum}{\end{eqnarray*}}
\newcommand{\banum}{\begin{eqnarray}}
\newcommand{\eanum}{\end{eqnarray}}
\newcommand{\bb}{\begin{thebibliography}}
\newcommand{\eb}{\end{thebibliography}}
\newcommand{\ci}[1]{\cite{#1}}

\newcommand{\bi}[1]{\bibitem{#1}}
\newcommand{\lab}[1]{\label{#1}}
\newcommand{\re}[1]{(\ref{#1})}
\newcommand{\Tr}{\mbox{Tr\,}}
\newcommand{\dd}{\partial}
\newcommand{\edc}{\end{document}}

\newcommand{\ga}{\disc{g_{A}}}

\newcommand{\mpi}{{\disc{m}_{\pi}}}

\newcommand{\gpinn}{\disc{g_{{\pi}NN}}}

\newcommand{\mn}{M_{N}}

\newcommand{\rhozero}{{\disc\rho_{0}}}

\newcommand{\ltwo}{{\cl}_{\mbox{2}}}

\newcommand{\bea}{\begin{eqnarray}}
\newcommand{\bc}{\begin{center}}
\newcommand{\ec}{\end{center}}
\newcommand{\eea}{\end{eqnarray}}

\newcommand{\lsk}{ {\cl}_{{\disc\mbox{sk}}} }

\newcommand{\dmuup}{\dd^{\mu}}
\newcommand{\dmud}{\dd_{\mu}}

\newcommand{\ds}{\displaystyle}
\newcommand{\dsf}{\displaystyle\frac}
\newcommand{\gsnn}{g_{\sigma NN}}
\newcommand{\gsnnstar}{g_{\sigma NN}^*}
\newcommand{\gastar}{\disc{g_A^{*}}}

\newcommand{\gpinnstar}{\disc{g_{\pi NN}^{*}}}

\def\theequation{\arabic{section}.\arabic{equation}}

\begin{document}
\draft

\title {\bf Density dependence of meson - nucleon vertices in nuclear matter.}
\author{A.M. Rakhimov$^{\dagger}$,
 F.C. Khanna$^{\ast}$,
U.T. Yakhshiev$^{\ddagger}$
and
 M.M. Musakhanov$^{\ddagger}$}

\address{
 $^{\dagger}$ Institute of Nuclear Physics, Academy of Science,
Uzbekistan (CIS)\\
 $^{\ddagger}$ Theoretical Physics Department,
Tashkent State University, Tashkent,\\
  Uzbekistan (CIS)\\
 $^{\ast}$ Physics Department, University of
 Alberta,\\
 Edmonton, Alberta T6G 2J1,\\
 and \\
 TRIUMF, 4004 Wesbrook Mall, Vancouver, BC V6T 2A3 }

\maketitle
\begin{abstract}
Changes in the meson-nucleon coupling constant and the vertex form
factor in nuclear matter are studied in a modified Skyrme Lagrangian
including the $\sigma$-meson field that satisfies the scale invariance.
Renormalization of $g_{A}$, the axial-vector coupling constant, and
the nucleon mass are studied in a consistent model. The results are
in good agreement with the empirical evidence. A calculation of the $\pi$-N
commutator, $\Sigma$-term, indicates that the medium changes its
magnitude considerably.
\end{abstract}

\vskip 5mm
\vskip 5mm


\newpage

\section{Introduction}

It is well  established that in - medium nucleon - nucleon (N-N)
cross sections manifest in heavy ion collisions
need not to be the same as their free space values \ci{cugnonalm}.
The origin of these changes must be reflected in modifications of
the N-N interaction which is predominantly one-boson exchange. The
parameters in meson exchange models are meson nucleon coupling
constants and their masses. To avoid divergencies in loop integrals
the meson - nucleon vertices are modified by form
- factors which effectively provide cut-off parameters. All these parameters
would change in the nuclear medium.

Even for the free N-N interaction the boson exchange models
use a phenomenological ansatz for vertex form factors.
Recently Holzwarth  and
Machleidt  \ci{holmach}  have shown that among the QCD inspired
pion - nucleon form factors the Skyrme form factor \ci{skff}
is the most preferable: it can describe well both the pion
 - nucleon and nucleon - nucleon systems.

The purpose of the paper is to investigate the role of the medium in
modifying the properties like meson nucleon coupling constant,
meson-nucleon form factors and the axial vector form factor and the
coupling constant. The calculations provide a consistent approach to
modification of these properties in a modified Skyrme model that may be
valid in a nuclear medium. The $\sigma$-meson is introduced as a
dilaton field to satisfy scale invariance. Even though it is well
known that scale invariance is badly broken, this provides a way to
introduce the $\sigma$-meson which is so essential for an N-N
interaction that is consistent with the two-body scattering data and
the bound deuteron.
In sections II and III the model will be presented; in
section IV the strong vertex form factors are derived and
in sections V and VI the renormalization of masses and coupling constants
respectively will be discussed. The summary and discussion of the
present approach are presented in the last section.

\section{The modified Skyrme Lagrangian}

In a previous paper  \ci{ourprc} a medium modified Skyrme Lagrangian
was proposed  and applied to study the static properties
of nucleons embedded in the nuclear medium. It gave a good description
of changes in nucleon mass and
its size in the medium. Here we shall outline the basic features of
the Langrangian and then extend it by including
the  dilated $\sigma$ meson field in order to satisfy the
scale invariance. It is well-known that scale invariance is badly
broken, this symmetry is retained here to investigate its consequences.

Our basic assumption in modifying of the Skyrme Lagrangian \ci{zbriska}
\be
{\cal L}_{\mbox{sk}}=\dsf{F_{\pi}^2}{16}
\Tr(\partial_{\mu}U)(\partial^{\mu}U^+)+
\frac{1}{32e^2}{\Tr}[U^+{\partial_{\mu}U}, U^{+}\partial_{\nu}U]^2-
\frac{F_{\pi}^2m_{\pi}^2}{16}{\Tr}(U^{+}-1)(U-1)
\lab{lsk}
\ee
where $F_{\pi}$ and $e$ are the parameters of the model, was that, in the
lowest order expansion
of the chiral field
$
U=\exp(2i\vec\tau\vec\pi /F_{\pi})\approx 1+2i
(\vec \tau\vec \pi)/{F_{\pi}}
-2{{\vec{\pi}}^2}/\fpi^2 +
\dots
$
the appropriate  medium modified Skyrme Lagrangian $\lsk^{*}$
 should give
the well known  equation for the pion field\ci{polar}:
\be
\partial_{\mu}\partial^{\mu}\pi+(m_{\pi}^2+\hat \Pi)\pi=0,
\lab{kg}
\ee
where $\mpi$ is the pion mass and $\hat \Pi$ is the self energy operator.
This may be achieved
simply by including $\hat \Pi$  in the pion
mass term of the Skyrme Lagrangian:\footnote{Here and below the
asterisk indicates the medium modified operators in quantities.}
\be
\lhisb^*=
-\dsf{F_{\pi}^2m_{\pi}^2}{16}{\Tr}[(U^+-1)(1+\hat \Pi/m_{\pi}^2)(U-1)].
\lab{lchisb}
\ee

In general,  the operator  $\hat \Pi=\hat \Pi_s+\hat \Pi_p$
in Eq.  \re{lchisb} acts both on
the center of mass coordinate, R,  and on the
internal coordinate, $\vec r$, of the Skyrmion. Only homogenous nuclear
matter is considered.
The translational invariance of the medium and that of the basic
Skyrme Lagrangian \re{lsk} makes the coordinate dependence of $\hat \Pi$
simpler:
$\hat \Pi\equiv \hat \Pi(\vec r-\vec R)$ which is also relevant
for a moving Skyrmion. Further we assume that the Skyrmion is placed
at the center of the nucleus i.e. $\vec R = 0$.

The $P$ - wave part of the pion self energy $\hat \Pi_{\Delta}$
is dominated by the
$P_{33}$ resonance. A simple model used
in practical calculations is the delta - hole model
which concentrates on the  pion - nucleon - delta
interaction ignoring the nucleon particle - hole excitations.
In momentum space the  $ \hat\Pi_{\Delta}$ is given by
$\hat \Pi_{p}(\omega,\vec k)=
\hat \Pi_{\Delta}(\omega,\vec k)=
-{\vec k^2\chi(\omega,\vec k)}/(1+\gzero\chi(\omega,\vec k))
$
\ci{polar} where $\gzero (= 0.6)$ is the Migdal parameter which
accounts for the short range correlations, also known as the
Ericson -  Ericson -  Lorentz -  Lorenz effect. The pion
susceptibility $\chi$  has nearly linear dependence on the
nuclear density $\rho$:
\be
\chi(\omega,\vec k)\approx\dsf{8}{9}\left(\dsf
{f_{\pi N\Delta}}{\mpi}\right)^2
\dsf{\rho\omega_{\Delta}}
{\omega_{\Delta}^2 -\omega^2}\exp(-2\vec k^2/b^2),
\ee
where $f_{\pi N\Delta}\approx 2 f_{\pi NN}$  ( $f^2_{\pi NN}/4\pi\approx
0.08$)
is the
coupling constant,
$\omega_{\Delta}\approx \vec k^2/2M_{\Delta}+M_{\Delta}-M_N$
and $b (\approx 7\mpi)$ is the range of the vertex form factor that
is chosen to have a gaussian form.
For the pions bound in the nuclear matter
$\omega$ is small $0\le\omega\le\mpi$, so
in the lowest order $\hat \Pi_{\Delta}$ has the form:
\be
\disc{\Pi_\Delta}(\omega,\vec{k})\approx - \vec k^2\chi_{\Delta}-\dsf{{\vec
k}^2\omega^2\alpha_{rt}}{\omega_{\Delta}^2},
\lab{polar}
\ee
where
$\chi_{\Delta}={4\pi c_0\rho}/{(1+4\pi\gzero\rho)},   $
$\alpha_{rt}={\chi_{\Delta}}/{(1+4\pi\gzero c_0\rho)}$  and
$
c_0={8}{f_{\pi N\Delta}^2}/{9m_{\pi}^2}{4\pi \omega_{\Delta}}.
$
In the static case $(\omega\rightarrow 0)$ it
coincides with the Kisslinger optical potential
\ci{polar}  which was used in the previous paper \ci{ourprc}.

Then in coordinate space the following symmetrized Lagrangian may be obtained
\be
\ba
\lsk^*=\dsf{\fpi^2}{16}\Tr \left(\dsf{\partial U}{\partial t}
\right)\left(\dsf{\partial U^+}{\partial t}\right)-\dsf{\fpi^2}{16}
(1-\chi_{\Delta})
\Tr(\vec \nabla U^+)(\vec \nabla U)+\\
\quad \\
+\dsf{\fpi^2}{16}\dsf{\alpha_{rt}}{\omega_{\Delta}^2}
\Tr\left(\dsf{\partial }{\partial t}
\dsf{\partial U^+ }{\partial r} \right)
\left(\dsf{\partial }{\partial t}
\dsf{\partial  U}{ \partial{r}}\right)
+\dsf{\fpi^{2} m_{\pi}^{*2}    }{16}   \Tr (U+U^+-2)+\\
\quad\\
+\dsf{1}{32e^2}{\Tr}[U^+{\partial_{\mu}U}, U^{+}\partial_{\nu}U]^2 ,
\lab{lfinalt}
\ea
\ee
where $m_{\pi}^{*2}=\mpi^2(1+\Pi_S(\rho)/\mpi^2)$, the effective pion
mass arises from the $S$ - wave part of the self energy
$\Pi_S(\rho) $.
Since the operators in nuclear matter do not satisfy the Lorentz invariance
there  are two different
effective coupling  constants  in the kinetic terms of
Eq. \re{lfinalt}\footnote {The similar second order
derivative terms had been suggested in ref \ci{wirsba}. }.
 The third term with "mixed" derivative in the Lagrangian
in Eq. \re {lfinalt} clearly vanishes in
 free space  ($\alpha_{rt}=0$  if $\rho=0$)
and contributes mainly to the moment of
inertia, $ I $, of the skyrmion. We shall consider this contribution
in section V by estimating the N$\Delta$ mass splitting.

\section {Inclusion of scalar meson}

Recently arguments have been given
in favor  of the empirical evidence for  a scalar-isoscalar meson, $\sigma$.
For example the isosinglet resonance  with a mass
$m_{\sigma}=553.3\pm 0.5$ MeV
and width $\Gamma=242.6\pm 1.2 MeV$
\ci{ishida} found in
the recent  $\pi \pi$  phase shift analyses is  believed to
have the properties
of the $ \sigma$ meson   that is essential for
the N-N potential in OBE  models to explain the NN-scattering data.

However  the $\sigma $  meson   as a chiral partner  of the pion
was originally  excluded from the Skyrme type  Lagrangians.
The only way of  including $\sigma$ meson  in the Lagrangian
is by  means of a dilaton field  appropriate to the scale transformations
$x\rightarrow \lambda^{-1} x$.  This may be achieved
in the following way: Consider any term in the Lagrangian and multiply it
by $\exp\{(d-4)\sigma\}$, with $d$ the number of derivatives. In fact,
since the scale invariance requires $\ds\delta\int dx^4 {\cal L}=0$,
each term of the Lagrangian should have  the scale  dimension
${\cal D}=4$. In this
sense the scale dimension ${\cal D}$ has values of $3/2$ and $1$ for
fermions and
bosons respectively. In the Skyrme
Lagrangian  the  scale  dimension of  the chiral field   equals
zero:    ${\cal D}[U]=0$  and  hence   ${\cal D}[\ltwo]=2$    and
${\cal D}[{\cal L}_4]=4$   by   itself.
As to the pion mass term, $\lhisb$, its  scale  property
is actually  uncertain for its origin may be
considered either  as a
quark - antiquark   pair  $\langle\bar{q}q\rangle$  with ${\cal D}=3$
  or a boson
 with ${\cal D}=1$. The former
is the   most  widely used   one  and this leads  to the introduction
of a   factor  $e^{-3\sigma}$   in the  pion  mass   term
i.e. $\lhisb\rightarrow\lhisb e^{-3\sigma}$.

The next  step is  the  choice of the self interaction term
$V({\sigma})$ of  these dilaton   fields. The  choice
$
V(\sigma)={e^{-2\sigma}\Gamma_0^2}(\dmud\sigma)(\dmuup\sigma)/2+
{C_g}[e^{-4\sigma}(1+4\sigma)-1]/4
$
is  commonly   used      and is uniquely   determined
by   the  trace anomaly    of  QCD \ci {gomm}.
However,  the scalar meson,   identified  with  fluctuations
of  the dilaton  field  associated  here  with the glueball,
is too  heavy  to be  considered  as a sigma meson mentioned above.
Its  inclusion in  the  chiral Lagrangian has small effects
on the nucleon    properties even at finite
densities \ci{meissmed}.
The following choice
\be
V(\sigma)=
\dsf{N_f\fpi^2}{16}e^{-2\sigma}(\dmud\sigma)(\dmuup\sigma)
+\dsf{N_fC_g}{48}\left[1-e^{-4\sigma}-\dsf{4}{\veps}
(1-e^{-\veps\sigma})\right],
\lab{vsig}
\ee
where $\veps=8N_f/(33-2N_f)$, $N_f$ is the number  of  flavors and
$C_g$ relates to the  gluon  condensate that  associates
the   dilaton  with  a quarkonium  which has  a reasonable
mass \ci{anov}. Although  the  dilaton   quarkonium
saturates  only  a part of the trace anomaly, namely the part which
is produced
by quark loops, but as part of the Skyrme
Lagrangian it gives  rather good description
of the nucleon static properties \ci{anov}   and the nucleon - nucleon
interactions  \ci{our}. The mixing angle
between the glueball and quarkonium fields  was found to be small \ci{anov}
and this is consistent with the QCD  sum rules.

Now a skyrmion imbedded  in nuclear matter is considered.
When a dilaton field is introduced into the effective
Lagrangian the dilaton potential $V({\sigma})$ would be
modified
by a $\sigma$ field generated by the medium itself.
While several attempts have been made \ci{kalfluid}, the correct nature of
this modification is still poorly
understood. On the other hand it is well known
that
the gluon condensate $C_g$ as well as the quark condensate
$\langle\bar{q}q\rangle$  decrease in the medium due to
the partial restoration of the
chiral symmetry. The present study is restricted
to considering the medium modification of
the dilaton potential (Eq. \re {vsig}) by taking into account
mainly the change of $C_g$ i.e. $C_g \rightarrow C_g^*$.

Thus putting all these considerations together the following
Lagrangian is proposed for homogeneous
nuclear medium in the static case
\be
\ba
\lsk^*=\dsf{\fpi^2\alpha_p\chi^2}{16}\Tr \vec L_i^2+
\dsf{1}{32e^2}{\Tr}[\vec L_i,\vec L_j]^2+
\dsf{\fpi^2m_{\pi}^{*2}}{16}\chi^3\Tr (U+U^+-2)-\\
\quad\\
-\dsf{\fpi^2}{8}(\vec\nabla\chi)^2+
\dsf{C_g^*}{24}\left[1-\chi^4-\dsf{4}{\veps}
(1-\chi^{\veps})\right],
\lab{lfinal}
\ea
\ee
where  $\chi(r)=e^{-\sigma(r)}$, $\alpha_p=1-\chi_{\Delta}$,
 $\vec L_i=U^+\partial_iU$.
Note that the Skyrme parameter $e$ coupled to the fourth
derivative term  remains unchanged
since this term is related to the exchange of a very heavy
$\rho$ - meson with mass
$m_{\rho}^*=m_{\rho}\rightarrow \infty$ \ci{bhaduri}.

There may be two alternative approaches to applying this Lagrangian in
nuclear physics.
In QHD like models, the $\sigma$ field plays
the role of an external field modifying the properties
of a soliton \ci{kalfluid} or a bag \ci{saitothomas} which
moves in the
background generated by the medium. In the
present model this approach would mean that $\sigma\equiv \sigma(R)$
and $U\equiv U(r)$ \ci{kalfluid}
that makes the scale invariance doubtful. In
contrast, the
mean field approximation is not used
for the $\sigma$ field. Instead the $\sigma$-field is strongly
coupled to the nonlinear pion fields so as to generate the soliton .
We assume that $\sigma\equiv\sigma(r-R)$ and  $U=U(r-R)$
are valid for a moving skyrmion.
In the previous paper \ci{ourprc} a similar approximation
(Eq. \re {lfinal} with $\sigma=0)$ was  used
to estimate the medium modified static properties of the nucleon and
found a well known behavior
of the  nucleon mass
$M_N^*/M_N<1$ and its size $R^{*}_{N} / R_{N} > 1 $.
In the next section we shall investigate in detail the dynamical
properties of the meson - nucleon system.

\section{Meson - nucleon form factors and Goldberger - Treiman relation}

The semiclassical procedure for calculating the meson - nucleon
vertex form - factors in a topological
chiral effective Lagrangian \ci{skff} is well-known.
In fact, the results of  more accurate methods
\ci{verchuehara}, based on the correct quantization
of the fluctuating chiral fields nearly coincide with the original
result that was given by Cohen \ci{skff}.
Using the ansatz $U(\vec r ,t)=A(t)U_0(\vec r-\vec R(t))A^+(t)$
 and defining the pion field as
\be
\pi_{\alpha}(\vec r)=-\dsf{iF_{\pi}}{4}\Tr [\tau_{\alpha}AU_0(\vec r-
\vec R)A^+]
\lab{quant}
\ee
the following expressions are obtained
\be
\ba
G^{*}_{\pi NN}(q)=\dsf{4\pi M_N^*F_{\pi}\alpha_p(\vec q^{\ 2}+
m_{\pi}^{*2}/\alpha_p)}{3q}\int\limits_0^{\infty} j_1(qr)
\sin(\Theta)r^2dr=\\
=\dsf{4\pi M_N^*F_{\pi}\alpha_p}{3}\int\limits_0^{\infty}
\dsf{j_1(qr)}{qr}S_{\pi}(r)r^3dr
\lab{ffpi}
\ea
\ee
for the pion nucleon form factor and
\be
\ba
G^{*}_{\sigma NN}(q)=2\pi F_{\pi}(\vec q^{\ 2}+m_{\sigma}^{*2})
\ds\int\limits_0^{\infty} j_0(qr)\sigma(r)r^2dr=
2\pi F_{\pi}\int\limits_0^{\infty}j_0(qr)S_{\sigma}(r)dr
\lab{ffsig}
\ea
\ee
for the sigma nucleon form factor respectively.
The details and the explicit expressions
for the source functions $S_{\pi}(r)$ and $S_{\sigma}(r)$
 are given in the Appendix.
Here we note that, in the chiral soliton models formulas
for  meson nucleon form -
factors are mainly determined by the quantization scheme
rather  than by  details of the Lagrangian.
The latter manifests itself through equations of motion
whose solutions $\Theta(r)$ and $\sigma(r)$ with spherically symmetric
 ansatz are
$U_0=\exp({i\vec{\tau}\vec{n}\Theta(r)})$, $\vec{n}=\vec{r}/r$, and
$\sigma(\vec r)\equiv\sigma(r)$
that should be used in Eqs \re {ffpi}, and \re {ffsig}.
The effective mass of the nucleon $M_N^*$
is given by
\be
M_N^*=M_H^*+\dsf{3}{8I^*}
\lab{mass}
\ee
where $M_H^*$ is the mass of the classical hedgehog soliton and
$I^*$ is the moment of inertia  of the spinning mode.

The effective
pion decay constant $f_{\pi}^*$ should be understood before
considering the Goldberger - Treiman (GT)
relation.
Medium renormalized pion - decay constant $f_{\pi}^*$ can be
naturally defined by
the PCAC relation:
\be
\vec \nabla \vec A_{\alpha}(x)=f_{\pi}^*m_{\pi}^{*2}\pi_{\alpha}(x)
\lab{pcac}
\ee
The medium modified axial coupling constant  $g_A^*$ measures
the spin - isospin correlations in a nucleon, embedded in a medium
and is defined as the expectation value of the space component
of the axial current $A^{i}_{\alpha} $
 in the nucleon state at zero momentum transfer
\ci{zbriska}:
\be
\lim_{q\rightarrow 0}\bra{N(\vec{P'})}A^{i}_{\alpha}(r)\ket{N(\vec{P})}=
\dsf{2}{3}\lim_{q\rightarrow 0}G^{*}_{A}(\vec{q}\ ^2)\bra{N}
\frac{\sigma_{i}\tau^{\alpha}}{2}\ket{N}\exp(i\vec{q}\vec{r})
\lab{limq}
\ee
where $\vec{q}=(\vec{P'} - \vec{P})$ and
$G_A^*(\vec{q}\ ^2)$ is the axial form factor of nucleon,
$G_A^*(0)=\gastar$. Here $\sigma_{i}$ is the component of the nucleon
spin. Due to the semiclassical quantization prescription, Eq.
\re {quant}, the matrix element of the pion field
evaluated between nucleon states is given by:
\be
\bra{N(P')}\pi_{\alpha}(\vec{r}-\vec{R})\ket{N(P)}=
\frac{\fpi\exp{(i\vec{q}\vec{r})}}{6}{\int}\bra{N}\sigma_{\alpha}
(\vec{\tau}\hat{x})\ket{N}e^{-i\vec{q}\vec{x}}\sin(\Theta)d\vec{x}
\lab{melpi}
\ee
Evaluation of the matrix elements between nucleon states
for both sides of Eq. \re {pcac} yields
\be
\gastar=\dsf{4\pi\fpi f_{\pi}^{*}m_{\pi}^{*2}}{9}
\int\limits_0^{\infty} \sin(\Theta)r^3dr
\lab{gasin}
\ee
that was originally derived in \ci{workman} for the free particle.
By comparing this equation with the expression for pion nucleon
coupling constant: $\gpinnstar=G^{*}_{{\pi}NN}(q^2)|_{q=0} $
 given by Eq. \re {ffpi}
the following  medium modified Goldberger - Treiman relation
is realized
\be
\disc{g_{\pi NN}^{*}}f_{\pi}^{*}=\gastar \mn^* \quad .
\lab{gt}
\ee
This relation has been proved \ci{meissmed,rho}.
On the other hand $\gastar$ and the axial form factor $G_A(q^2) $
may be calculated directly from the Lagrangian in Eq. \re {lfinal}
in terms of the Noether currents
\ci{zbriska}. This gives
$$
G_A(q^2)=
-4\pi\int\limits_0^{\infty}[j_0(qr)A_1(r)+\dsf{j_1(qr)}{qr}A_2(r)]r^2dr,
$$
with
\be
\ba
A_{1}(r)=\dsf{s_2}{8r}\left[e^{-2\sigma}\alpha_pF_{\pi}^2+
\dsf{4}{e^2}\left(\Theta^{\prime 2}+d\right)\right],\\
\quad \\
A_{2}(r)=-A_{1}(r)+\dsf{\Theta^{\prime}}{4}\left(\fpi^2\alpha_p
e^{-2\sigma}+\dsf{8d}{e^2}\right)
\lab{gaff}
\ea
\ee
and
\be
g_A^*=-\dsf{\pi}{3}
\int\limits_0^{\infty}\left\{e^{-2\sigma(r)}F_{\pi}^2\alpha_p
\left(\Theta^{\prime}+\dsf{s_2}{r}\right)+\dsf{4}{e^2}
\left[(\Theta^{\prime 2}+d)\dsf{s_2}{r}+
2\Theta^{\prime}d\right]\right\} r^2dr
\lab{ga}
\ee
where $d=\sin^2(\Theta)/r^2$, $s_2=\sin(2\Theta)$, and $\alpha_p$
 is defined in Eq. \re {lfinal}.
The renormalized pion decay constant $f_{\pi}^*$ is obtained by
combining the results given in Eq.s \re{gt} and \re{ga}.

\section{ Renormalization of hadron masses}

Before going to a quantitative analyses of the medium effects
we fix the following set of parameters for free space:
$F_{\pi}=186MeV$, $m_{\pi}=139MeV$, $C_g=(260MeV)^4$.
 This gives a good description  of the
sigma meson properties $m_{\sigma}=550MeV$, $\Gamma_{\sigma}=251.2MeV$,
that may be compared with their
experimental values obtained from the recent $\pi\pi$
phase shift analysis \ci{ishida}.
The Skyrme parameter $e$ has been adjusted to reproduce the pion nucleon
coupling constant: $g_{\pi NN}=13.5$ for $ e=4.05$.
The well-established fixed parameters of the $P$ - wave pion self
energy in  Eq. \re {polar} are used in the pion
sector: $\gzero=0.6$, $c_0=0.13m_{\pi}^{-3}$ \ci{ourprc,polar}.

Now, the medium dependence of the input parameters may be considered.
The possible
renormalization of the Skyrme parameter $e$ cannot be
studied in the present approach unless the $\rho$ meson
is included in the Lagrangian explicitly. So we take
$e^*=e$.

We adopt the following parametrization of $m_{\pi}^*$
\be
m_{\pi}^*=m_{\pi}\sqrt{1+\hat{\Pi}_{s}(\rho)/m_{\pi}^{2} }=
m_{\pi}\sqrt{ 1-4{\pi}b_{0}\rho\eta/m_{\pi}^{2}} \ ,
\lab{mpi}
\ee
where $\eta=1+\mpi/\mn $ and $b_0$ is an effective
S - wave $\pi - N$ scattering length. It is anticipated \ci{ourprc}
that the results will not be
to sensitive to the value of $b_0$.

The only  input parameter in the scalar meson sector is
$C_g^*$. The medium renormalization of the gluon condensate
$C_g^*$,
in contrast with the renormalization
of the quark condensate $\langle\bar{q}q\rangle$
 \ci{birserev} and meson masses \ci{ko},
is poorly known.
However in the present approach \re{lfinal}
$C_g^*$  may be determined  by
$m_{\sigma}^*$  through the equation
\be
 C_g^*=\dsf{ 3\fpi^2m_{\sigma}^{*2} N_f}{4(4-\veps)}.
\lab{cg}
\ee
Various
approaches \ci{meissmed,saitothomas,rho}  show that $m_{\sigma}^*$, has
a linear density dependence. The following parametrization
\be
\dsf{m_{\sigma}^*}{m_{\sigma}}=1-0.12\dsf{\rho}{\rho_0}
\lab{msigma}
\ee
is adopted here. It is consistent with the one obtained
in the QCM framework \ci{saitothomas}.

The results for the static properties of hadrons are presented
in Table I. The  second ($m_{\pi}^*$) and the third $(m_{\sigma}^*)$
columns  of  the table
should be considered as  input data for they were taken from
other models \ci{polar,saitothomas}. The medium renormalized
gluon condensate $C_g^*$ is calculated from Eq.s \re {cg}
and \re{msigma}
with $N_f=2$.
The change in the gluon
condensate is small $\sim 5\%$  at  normal nuclear matter
density, $\rho_0=0.5m_{\pi}^3$.
The stiffness of the gluon condensate  as a consequence of
the lack of scale invariance of QCD has been shown
by Cohen \ci{cohencg} who found that the fourth
root of the condensate might be
altered by no more than $4\%$ \ci{birserev}.

The main contribution to the $\sigma\pi\pi$ vertex,
 and  hence, to the decay width of $\sigma $ meson at the tree level:
$
\Gamma_{\sigma\rightarrow \pi\pi}={m_{\pi}^{*3}x^3\alpha_p^2
\sqrt{1-4x^2}(1-2x^2)^2}/{4\pi F_{\pi}^2},
$
where $x=m_{\pi}^*/m_{\sigma}^*$,
arises from the first term of the Lagrangian in Eq. \re {lfinal}.
The table shows that the  width
 $\Gamma_{\sigma\rightarrow \pi\pi}$ is decreased significantly
in the medium. This stimulates an interest to observe the
$\sigma$ mesons in nuclei by experiments
proposed earlier \ci{kunihiro}.

Now medium effects on the mass of nucleon
$M_N^*$ and $N\Delta$ mass splitting $\delta M_{N\Delta}=
M_{\Delta}^*-M_N^*$ will be considered. In general, it is almost impossible
to reproduce simultaneously the experimental values of
 masses and coupling
constants within the Skyrme model even for a free particle.
Since dynamics is the main interest, the set of parameters
was chosen so as to
reproduce the pion nucleon coupling constant: $\gpinn=13.5$.
It is clear from Table I that the free space value of the nucleon
mass $M_N$ is slightly large $\mn=1413 MeV$, whereas $\delta M_{N\Delta}$
 is reproduced
rather well $\delta M_{N\Delta}=284MeV$
($\delta M_{N\Delta}^{\mbox{exp}}=293 MeV)$.
 The effective mass of the nucleon $\mn^*$
 in normal nuclear matter density is
decreased by a factor of $ \mn^*/\mn=0.82  $ which is
in a good agreement with the
estimates based on QCD sum rules $M_N^*(QCD)=680\pm 80MeV$  i.e.
 $\mn^*/\mn=0.72\pm0.09 $ \ci{furnstahl}.
We underline that the mass of the nucleon
should be treated
as the mass of the baryon which  emerges
as a soliton in the  sector with baryon  number one (B=1).

The study of $N\Delta $ mass splitting gives  a chance to estimate the
contribution from the "mixed derivative" term - the third term
on the r.h.s of Eq. \re {lfinalt}:
\be
{\cal L}_{rt}=
\dsf{\fpi^2}{16}\dsf{\alpha_{rt}}{\omega_{\Delta}^2}
\Tr\left(\dsf{\partial }{\partial t}
\dsf{\partial U^+ }{\partial r} \right)
\left(\dsf{\partial }{\partial t}
\dsf{\partial  U}{ \partial{r}}\right)
\lab{lrt}.
\ee

 Actually, owing to the canonical quantization
$\delta M_{N\Delta} $ is related to the moment
of inertia $I$ by:
$
\delta M_{N\Delta}=M_{\Delta}^*- M_{N}^*=3/2(I^{*}_{0}+I_{rt}),
$
where  $I_{rt}$  is the net contribution from ${\cal{L}}_{rt}$
(clearly $I_{rt}=0$ in free space). The explicit expressions for
 $I^{*}_{0}$ and $I_{rt}$ are given in the Appendix.
In Table I are shown  $\delta M_{N\Delta}$ that
has been   calculated  with the inclusion of $I_{rt}$
(denoted here $\delta M_{N\Delta}^a$) and without the inclusion of
$I_{rt}(\delta M_{N\Delta}^b)$.
In the nuclear medium the ${\cal{L}}_{rt}$
term
leads to an enhancement of  the moment of inertia  decreasing
$\delta M_{N\Delta}$ significantly. Even without the term
${\cal{L}}_{rt}$ in the lagrangian
the shift of $\delta M_{N\Delta}$
from its free value
$\delta M_{N\Delta}^*/\delta M_{N\Delta}=0.75$
is larger
than that obtained by
Meissner \ci{meissmed} in a medium modified chiral soliton
model based on the Brown Rho (BR) scaling law
($\delta M_{N\Delta}^*/\delta M_{N\Delta}=0.87)$.

\section{Renormalization of coupling constants and form - factors}
\setcounter{equation}{0}

The axial - vector exchange currents
 must be considered in order to investigate the medium
effects on $g_A$ and on the axial form factor $G_A(q^2)$. However it
is known that the bulk of exchange current effects arise from the
$\triangle$-hole contributions. Including such $\triangle$-h effects
would imply that bulk of the exchange currents effects are included
in the effective $g_{A}$. In a
heavy nucleus it is meaningful to take these
axial exchange operators into account as
corrections to the \underline{effective} axial current operator
of a single nucleon $\vec A_{\alpha}=-g_A^*\vec \sigma\vec\tau_{\alpha}$.
So $g_A^*$ in Eq. \re {ga}
may be considered as an effective axial coupling
constant modified by the medium polarization and screening effects
since we
are considering an effective
one body problem of a nucleon embedded in the nuclear
medium.

The second column of Table II displays a well
known quenching behavior of $\ga$ that is  mainly caused by a
  factor $\alpha_p=1-\chi_{\Delta} < 1 $ in the
first term of Eq. \re {lfinal}.
Note that the same set of input parameters $\fpi, e, c_0$ (but without
dilaton field $\sigma=0) $ gives the desired ratio $\gastar/\ga=0.8 $
\ci{ourprc}. The present calculations show that the inclusion of
the scalar meson, which induces an additional attraction,  prevents
larger quenching: $\gastar/\ga=0.9 $.

In nuclei, a nucleon polarizes the medium in its vicinity.
This leads to a screening effect that reduces the effective
pion nucleon coupling strength.
In the present approach the screening mechanism may be described
as being due to virtual $\Delta h$ excitations
that have been taken into account
by the self - energy $\disc{{\hat\Pi}_\Delta}$ term in Eq. \re {polar}.
At normal  nuclear matter density the renormalization
of $g_{\pi NN}$ amounts to a reduction of $25\%$ of
the coupling strength. This is sufficient to explain the quenching of
the Gamov - Teller strength
in heavy nuclei \ci{khannatowner}. Furthermore this is consistent
with a general argument based on Ward-Takahashi relations\ci{xzh}.


The effective pion
decay constant $f_{\pi}^*$  is obtained by using
the GT relation  \re {gt}. Comparing the ratios
$f^{*}_{\pi}/f_{\pi}$ (Table II) and $M_N^*/\mn$ (Table I)
one may see that they both decrease in the nuclear medium.
This fact is in good qualitative agreement with
Brown-Rho scaling law
\ci{br} predicted within a simple Skyrme model.

The masses of $w$- and $\sigma$-meson are supposed to decrease by the
scaling law. There are experimental indications that this is true. In
the same way the $\sigma$N coupling constant is expected to decrease.
The ratio of $\gsnnstar$ to its free value  is presented in Table II.
The changes
in $\gpinn$ and $\gsnn$ are nearly the same. Both are reduced
in the medium by $\approx 25\%$ at $\rho=\rhozero$.

In Figs 1 and 2, the renormalized
 $\pi NN$  and  $\sigma NN$ vertex
form factors respectively
at $\rho=\rho_0$ (dashed lines) in comparison with
these in the free space (solid lines) are displayed. Appreciable
quenching of both form factors is observed. At small momentum transfer
these can be parametrized by a monopole form i.e.
$G_{\pi NN}(\vec q\ ^2)=\gpinn/(1+\vec q\ ^2/\Lambda_{\pi}^2)$ and
$G_{\sigma NN}(\vec q\ ^2)=\gsnn/(1+\vec q\ ^2/\Lambda_{\sigma}^2)$.
Table II shows that the cut off parameter $\Lambda_{\pi}$ is decreased
significantly
at $\rho=\rho_0$. Relatively
small changes in $G_{\sigma NN}(\vec q\ ^2)$ seem to be caused by a
stiffness of the $ \sigma $-field or equivalently
  $C_g^*$  \ci{birserev}.

The nucleon axial form factor
$G_{A}(\vec{q}^{\ 2})/G_{A}(0)$calculated for $\rho=0$ and
 $\rho=\rhozero$  is presented in Fig. 3 with solid and
dashed curves respectively. It is seen that, the modification of
$G_{A}(\vec{q}^{\ 2})$ is not as simple as that of
$G_{\pi NN}(\vec{q}^{\ 2})$.
The medium leads to a quenching of the meson nucleon
form factors  over a range of $\vec{q}^{\ 2}$,
while the quenching of $G_{A}(\vec{q}^{\ 2})$ takes place
at $\vec{q}^{\ 2} =0$ and $\vec{\ q}^{2} > 10 fm^{-2}$.

A further interesting quantity is
the pion-nucleon sigma term $\Sigma_{{\pi}N}$.
It is both the chiral symmetry breaking piece of the nucleon mass and a
measure of the scalar density of quarks inside the nucleon.
Due to the Hellman
Feynman theorem,
 it is easily calculated \ci{hftheorem} in the Skyrme model
\be
\Sigma_{\pi{N}}=m_{q}\frac{\partial{M_N}}{\partial{m_q}}=
\frac{\partial{M_N}}{\partial{m_{\pi}^{2}}}m_{\pi}^{2}  .
\lab{sigman}
\ee
For free space ($\rho=0$) the lagrangian including scalar mesons
Eq.\re{lfinal}
 gives $\Sigma_{\pi{N}}\approx 20.1 MeV$ that is much smaller than that
obtained in the original skyrme model \ci{zbriska}. It may be argued
that $\Sigma_{\pi{N}}$ may also undergo changes in the nuclear
medium i.e.  $\Sigma_{\pi{N}}^*\ne\Sigma_{\pi{N}}$.
Actually, PCAC allows us to relate $\Sigma_{\pi{N}}$ to the soft-pion
limit of $\pi{N}$ scattering \ci{reua}
whose parameters may  not be the same in
free space and the medium.
 So, defining an effective value of the in-medium
pion-nucleon sigma commutator as
\be
\Sigma_{\pi{N}}^*=m \  {}^*\!\bra{N}{\ds\int}d\vecr\bar\psi\psi\ket{N}^*=
\setcounter{footnote}{0}\frac{\partial{M_N}^*}{
\partial{m_{\pi}^{*2}}}m_{\pi}^{*2}
\lab{sigmanstar}
\ee
where $\ket{N}^*$  is the state
of the nucleon bound in
nuclear matter, we obtain $\Sigma_{\pi{N}}^*\approx 40.1 MeV$ at normal
nuclear
density. This means a large increase of the nucleon sigma term:
$\Sigma_{\pi{N}}^*/\Sigma_{\pi{N}}\approx 2$.
Using appropriate solutions of Eqs. \re{Eq}, it is
estimated that
\be
\dsf{\Sigma_{\pi{N}}^*}{\Sigma_{\pi{N}}}{\approx}
\dsf{m_{\pi}^{*2}
\ds\int\limits_0^{\infty}e^{-3\sigma^*}[1-\cos(\Theta^*)]x^2dx}
{m_{\pi}^{2} \ds\int\limits_0^{\infty}e^{-3\sigma}[1-\cos(\Theta)]x^2dx}=
1.88\dsf {m_{\pi}^{*2} }{m_{\pi}^{2} } \ ,
\ee
where $\sigma^*(x) $,  $ \Theta^*(x)$ are profile functions
at $ \rho=\rhozero$ and  $\sigma(x) $, $ \Theta(x)$
are those at $\rho=0$.
Thus,  the medium renormalization of
$\sum_{\pi{N}}$ is caused mainly due to
a large modification of the
profile functions (see Figs. 4a, 4b).
On the other hand $\Sigma_{\pi{N}}$ gives a good estimate for the
quark condensate at finite density:
\be
\frac{\langle\bar{q}q\rangle_{\rho}}{\langle\bar{q}q\rangle_{vac} }=1-
\frac{\Sigma_{\pi{N}}^*\rho}{\mpi^{2}f_{\pi}^{2}}
\ee
Hence, assuming the last equation  holds it is concluded that the
in medium enhancement of $\Sigma_{\pi{N}}$
 leads to further quenching
of the scalar density of quarks $\langle\bar{q}q\rangle_{\rho}$ in nuclear
matter.

\section{Discussions and summary}
\setcounter{equation}{0}

We have proposed a medium modified Skyrme like Lagrangian
which takes into account the distortion of the basic nonlinear
meson fields by the nuclear medium.
It is extended by including the scalar - isoscalar
sigma meson which is identified with a dilaton - quarkonium.
 The influence of the medium on pion
fields is introduced  by the self energy
operators $\hat\Pi_{p},(=\hat\Pi_{\Delta})$ and $\hat\Pi_{s}$  while the
effect of the  medium on the
 dilaton field
is limited to the renormalization of the gluon condensate. The Lagrangian
is
applied to study changes in the hadron masses and meson - nucleon
vertex form - factors.

In particular,  the mass of the $\Delta$ - resonance decreases
more than that of the nucleon in the nuclear medium. Consequently the pion
requires
lesser energy to excite the nucleon to the $\Delta$
in the nuclear medium than it does for a free nucleon.
The mass difference between $\Delta$ and N decreases to $42\%$ of that
for free particles at the nuclear matter density. This is quite
consistent with earlier estimates \ci{akira,meissmed} but contradicts
the recent  theoretical results of Mukhopadhyay and Vento \ci{vento},
who found $\delta M_{N\Delta}^{*}/\delta M_{N\Delta}\approx 1.25$
for $ \rho=0.8\rhozero$.
So, it would be quite interesting to study N$\Delta$ mass splitting
experimentally by an analyses of the N$\Delta$ transitions in heavy nuclei.

We have investigated the characteristic changes of the decay width
$\Gamma_{\sigma\rightarrow \pi\pi}$
at zero temperature. One may expect that the temperature-dependence of
the physical
quantities is qualitatively similar to the $\rho$ dependence.
 In this sense, our results are in good agreement
with predictions of the in - medium   NJL model \ci{hatsudakunihiro},
that at sufficiently high temperatures the $ \sigma$ meson becomes a
sharp resonance and its width may even vanish.
Clearly, more precise predictions on
$\Gamma_{\sigma\rightarrow\pi\pi} (T,\rho)$ in the framework
of the present Lagrangian should be made by studying thermal Green's
function of $\sigma$ - meson e.g. within  Thermo Field Dynamics.

Furthermore the in - medium version of GT relation
\re{gt} holds in the present approach. The renormalized pion
decay constant $f_{\pi}^{*}$ and nucleon mass $M_N^{*}$ do not satisfy
the BR scaling \ci{rho}: the change in the nucleon mass is larger
than that in $f_{\pi}$.

The medium effects lead to a quenching of the meson - nucleon form factors
as well as the coupling constants   $g_{\pi NN} $ and $ g_{\sigma NN}$.
  The latter  should be
compared with the results of Banerjee and Tjon \ci{banerjeetjon}
obtained in the framework of CCM.
There $\gsnn$ and $\gpinn$ increased at low densities
while in the present approach and in QCM \ci{saitothomas}
they decrease with density.

What is the possible origin of this
descrepancy.Possible in - medium changes in nucleon properties
and in meson - nucleon dynamics should be understood
within those effective field thories where a nucleon has an internal structure.
 In general the models on which they are based differ
even at the  single nucleon level.Moreover these models may
 be radically different from one another at a deeper level
such as the confinement mechanism for quarks or a realization of the chiral
invariance.
For example, one may easily find that there is no confinement in the
Skyrme model by construction while it is implemented in CCM
\ci{banerjeetjon} more carefully than in QCM. Now it is natural to propose
that these differences are reflected not only in the free space
but also in the case when the nucleon, and hence
quarks,  are  influenced  by the medium. It maybe concluded
 that all "starred"
features of nucleon (mass, copuling constants etc) are model
dependent and, comparing the present model with the CCM model of
Banerjee and Tjon or QCM may not be entirely straight forward.

 However we believe that
the natural reduction of the meson  masses and coupling strengths
found in the present model
are expected to give a good description
of the saturation properties of nuclear matter.
It would be interesting to calculate properties of finite nuclei and
nuclear matter at normal and high density to establish the validity
of the model. The heavy ion reactions would possibly open a new window
at finite temperature properties of nucleons in nuclei and other
nuclear properties. In particular a study of the change of the effective
mass and coupling constants with temperature would shed new light on
the type of effective Lagrangian that is appropriate for the system.

By introducing a formal definition of the in - medium  pion - nucleon
sigma term\\
$\Sigma_{\pi{N}}^*=m_{\pi}^{*2}
[{\partial{M_N}^*}/{\partial{m_{\pi}^{*2}}}]$,
it is  found that in contrast to the meson - nucleon coupling strengths
the      $\Sigma_{\pi{N}} $ increased in the medium:
$\Sigma_{\pi{N}}^* / \Sigma_{\pi{N}}\approx 2 $. This enhancement
could lead to a decrease of the quark condensate
$\langle{\bar q}q\rangle_{\rho}$  in nuclear matter.
However as it was recently pointed out
by Birse \ci{birse96} this change should not lead to a
drastic and rapid restoration of the chiral symmetry in nuclear matter.

It is anticipated that
forthcoming ultrarelativistic heavy - ion collision experiments (e.g. at RHIC)
will provide  significant new information on the strong interactions
through the detection of changes in hadronic properties \ci{quarkmatter}.
This would provide an impetus to consider refined models for strong
interactions in nuclear matter at high density and at finite
temperature.

It is useful to comment on the role of the scale invariance
in nuclear physics and its breaking by the trace anomaly.
 The physical scalar - isoscalar
sigma meson has been considered here as a dilaton - quarkonium
rather than a heavy quarkonium which is used in standard approaches
\ci{gomm}.In a more general case a heavy scalar gluonium
\underline{and} light scalar quarkonium are both needed to saturate
the low energy theorems involving the trace anomaly
 of energy- momentum tensor. Fortunately, it was shown
\ci{furnrho} that the heavy gluonium degree of freedom can be integrated out,
so that the contribution from the light quarkonium becomes effectively
larger (i.e. dominates). In this sense the approach given by eq. \re{vsig}
is similiar
to the one used in \ci{furnrho} \footnote{We are indebted to M. Rho
for pointing this to us}, although in the latter
case the dilaton quarkonium has an anomalous scale dimension
i.e.\\
$$ {\cal D}_{\mbox{an}}={\cal D}_{\sigma} - 1 {\ne} 0.$$
 Actual calculations
in the framework of a chiral effective lagrangian for nuclei
with structureless nucleons predicted rather large value
 :${\cal D}_{\mbox{an}}\approx 5/3$, or in general
 ${\cal D}_{\mbox{an}}   >1$,
otherwise bulk properties of finite nuclei cannot be reproduced.
On the other hand, one may ask here, how a model with such
a large anomalous dimension of scalar quarkonium
would describe properties of a single nucleon in free
space? This question may be answered by including the
dilaton quarkonium with  ${\cal D}_{\mbox{an}} {\ne}0 $ into the Skyrme model
or into the chiral soliton model \ci{meisrep} by means
of scale invariance and trace anomaly. This work is in progress.

In conclusion the modified Skyrme Lagrangian with scale invariance
provides a useful insight into the role of medium in changing various
properties of the mesons and nucleons. Even though scale invariance
is badly broken in strong interactions its inclusion gives important
information on the role of this symmetry property in a many-particle
system.

\begin{center}
{\bf ACKNOWLEDGMENTS}
\end {center}

We would like to thank Yu. Petrov, W. Alberico and K. Fujii
for helpful comments on the issues
discussed here during the Conference "Structure of Particles, Nuclei
and their interactions" held in Tashkent. The research of
F.C. Khanna is supported in part by the Natural Sciences and Engineering
research Council of Canada. The research of U.T. Yakhshiev and
M.M.Musakhanov is supported in part by grant N11/97 GKNT RUz. We
thank M. Rho and M.K. Banerjee for useful comments on an earlier
draft.

\section*{appendix}
\def\theequation{A.\arabic{equation}}
\setcounter{equation}{0}

The expressions for $G_{\pi NN}^{*}(q^{2})$ and $G_{\pi NN}^{*}(q^{2})$
are derived following the method given by Cohen \ci{skff}.
The small fluctuations around the vacuum value are related to the pion
field by
$
U=\exp(2i\vec\tau\vec\pi /F_{\pi})\approx 1+2i
\vec \tau\vec \pi/{F_{\pi}}-2{{\vec{\pi}}^2}/\fpi^2+\dots
$
which gives the following approximation for the Lagrangian
in Eq. \re{lfinal}:
${\cal L}\approx -\dsf{1}{2}(\vec \nabla \vec \pi)^2\alpha_p-
\dsf{1}{2}m_{\pi}^{*2}\vec \pi^2$ and for the equation of motion:
 $ -(\vec \nabla^2 \pi)\alpha_p+m_{\pi}^{*2}\vec \pi=0$.
The factor $\alpha_p=1-\chi_{\Delta}$ in the last equations is
obtained by
renormalization of the pion propagator in the medium i.e. by the $\Delta$
- hole self energy term $\hat\Pi_{\Delta}$.
We define the in-medium $\pi NN$ coupling constant and
the vertex form factor by introducing the source term
$\vec j=iG_{\pi NN}^*\bar \psi\vec \tau\gamma_5\psi$ into the
equation of motion:
\be
-\nabla^2\vec \pi\alpha_p+m_{\pi}^{*2}\vec \pi=iG_{\pi NN}^*
\bar \psi\gamma_5\vec \tau\psi ,
\lab{appkg}
\ee
where
\begin{eqnarray}
\psi\ket{N(\vec P)}=\dsf{e^{i\vec P\vec x}}{(2\pi)^{3/2}}
\left(\dsf{E^*+M_N^*}{2E^*}\right)^{1/2}
\left(\begin{array}{c}
1\\
(\vec \sigma\vec P)/(E^*+M_N^*)
\lab{psi}
\end{array}\right)\chi_S,
\end{eqnarray}
where $E^{*2}=\vec P^2+M^{*2}$.
In the Breit frame $(\vec P '=-\vec P$,
 $\vec q=\vec P -\vec P')$ the matrix element of the source
evaluated between nucleon states is given by:
\be
\bra{N(\vec P')}\vec j(r)\ket{N(\vec P)}=
\dsf{e^{i\vec q\,\vec r}\bra{N}(\vec \sigma\vec q)\ {\vec \tau}\ket{N}}
{(2\pi)^{3} \ \ 2M_N^*}.
\lab{appmelj}
\ee
Using the quantization rules
\be
\ba
\pi_{\alpha}(r)=-\dsf{iF_{\pi}}{4}\Tr[\tau_{\alpha}AU_0(r-R)A^+],\\
\quad \\
\bra{N}\Tr\tau_{\alpha}A\tau_{\beta}A^+\ket{N}=
-\dsf{2}{3}\bra{N}\sigma_{\alpha}\tau_{\beta}\ket{N}
\ea
\ee
we have
\bea\ba
\bra{N'(\vec P')}\pi_{\alpha}(\vec r,\vec R)\ket{N(\vec P)}=
\dsf{e^{i\vec q\,\vec r}}{(2\pi)^{3}}\ds\int
\bra{N'}\pi_{\alpha}(x)\ket{N}e^{-i\vec q\,\vec x}d\vec x=\\
\quad \\
=\dsf{F_{\pi}e^{i\vec q\,\vec r}}{6(2\pi)^{3}}\ds\int
\bra{N'}\sigma_{\alpha}(\vec \tau\vec x)\ket{N}
e^{-i\vec q\,\vec x}\sin(\Theta)d\vec x
\lab{appmelpi}
\ea\eea
where $\Theta$ is defined
by the hedgehog ansatz $U_0(r)=e^{i(\vec \tau\vec r)\Theta(r)}$.
Now the matrix element of Eq. \re{appkg} is evaluated between collective
wave functions
for spin-up proton $|p\uparrow\rangle$ with momentum $\vec P$ and
 $\vec P'$ in the Breit frame
and using equations \re{psi} - \re{appmelpi}
to obtain
\bea\ba
G_{\pi NN}^*(\vec{q}^{\ 2})=-\dsf{iF_{\pi}\alpha_pM_N^*}{3q}\int
(-\nabla^2+m_{\pi}^{*2}/\alpha_p)\hat x_3\sin(\Theta)
e^{-i\vec q\,\vec x}d\vec x=\\
\quad \\
=\dsf{4\pi F_{\pi}\alpha_pM_N^*}{3}\ds\int\limits_o^{\infty}
\dsf{j_1(qx)}{qx}S_{\pi}(x)x^3dx
\ea\eea
where $S_{\pi}(x)=-2\Theta'c/x-\Theta''c+\Theta's+2s/x^2+
m_{\pi}^{*2}s/\alpha_p$ with $c=\cos(\Theta)$, $s=\sin(\Theta)$.

Similarly, the coupling constant at the $\sigma NN$
vertex is defined by the equation:\\
$(-\vec \nabla^2 +m_{\sigma}^{*2})\sigma=G_{\sigma NN}^{*}\bar\psi\psi$.
Evaluating matrix elements of both sides of this equation
it is easy to obtain the $\sigma NN$ form factor:
\bea\ba
G_{\sigma NN}^*(\vec{q}^{\ 2})=2\pi F_{\pi}
\ds\int\limits_o^{\infty}  j_0(qx)S_{\sigma}(x)dx\    ,
\ea\eea
with $S_{\sigma}(x)=-x^2\sigma''+2x\sigma'+x^2m_{\sigma}^{*2}\sigma$.

Note that the profile functions $\Theta(r)$ and $\sigma(r)$
in $S_{\pi}(x)$ and  $S_{\sigma}(x)$ are the solutions of the equations
of motion:
\bea\ba
\Theta''x^2\chi^2\alpha_p+4s_2\Theta^{\prime 2}+8s^2\Theta^{\prime\prime}+
2x\Theta'\chi^2\alpha_p+2x^2\Theta'\chi\chi'\alpha_p-\\
\qquad\qquad
-\chi^2\alpha_ps_2-4s_2d-x^2\beta^2\chi^3s=0\\
x^2\chi''+2\chi\chi'-2\chi\alpha_{p}x^2(\Theta^{\prime 2}/2+d)-
16x^2{\cal D}_{eff}(\chi^3-\chi^{\varepsilon-1})-\\
\qquad\qquad
-3x^2\beta^2(1-c)\chi^2=0.
\lab{Eq}
\ea\eea
where $x\equiv eF_{\pi}r$, $s_2\equiv\sin (2\Theta)$,
$d\equiv s^2/x^2$, $\beta=m_{\pi}^*/e F_{\pi}$,
${\cal D}_{eff}=C_g^*/24e^2F_{\pi}^4$,\\ $\chi\equiv\exp(-\sigma(x))$.
The boundary conditions are:\\
 $\Theta=\pi$, $\sigma'=0$ for $x\rightarrow 0$ and
$\Theta \sim (1+\beta x)e^{-\beta x}/x^2$, $\sigma\sim \Theta^2$
for large $x$.

For completeness we write also the explicit expressions for
$I=I_0^*+I_{rt}$:
$$
I_0^*=\dsf{2\pi}{3e^3F_{\pi}}\int\limits_0^{\infty} s^2\{e^{-2\sigma}+
4(\Theta^{\prime 2}+d)\}x^2dx,\qquad
I_{rt}=\dsf{\pi\alpha_{rt}}{3\omega_{\Delta}^2eF_{\pi}}
\int\limits_0^{\infty} \{\Theta^{\prime 2}c^2+2d\}x^2dx.
$$

\bb{99}
\bi{cugnonalm}
   J. Cugnon, A. Lejeunne and P.Grande, Phys. Rev. C 35 (1987) 861;
   T. Alm, G. Ropke and M. Schmidt, Phys. Rev. C 50 (1994) 931.
\bi{holmach}
   G. Holzwarth and R. Machelidt, Phys. Rev. C 55 (1997) 1088.
\bi{skff}
   T.D. Cohen, Phys. Rev. D 34 (1986) 2187;
   N. Kaiser, U. Vogl, W. Weise and U. G. Meissner,
   Nucl. Phys. A 484 (1988) 593.
\bi{ourprc}
   A.M. Rakhimov, M.M. Musakhanov, F.C. Khanna and  U.T. Yakhshiev,
        nucl - th/9609049  (submitted to \prc).
\bi{zbriska}
   G.S. Adkins and C.R. Nappi, Nucl. Phys. B 233 (1984) 109;
   I. Zahed and G.E.Brown,  Phys. Rep.  142 (1986) 1;
   E.M. Nyman and D. O. Riska, Int. J. Mod. Phys. A 3 (1988) 1535.
\bi{polar}
   T. Ericson and W. Weise, Pions and nuclei, (Claredon-Press,
        Oxford, 1988).
\bi{wirsba}
   A. Wirsba and V. Thorsson, "In - medium effective chiral Lagrangians
        and pion mass in nuclear matter", hep-ph/9502314.
\bi{ishida}
   S. Ishida et al., Progr. Theor. Phys. 95 (1996) 745;
   M.Y. Ishida,  Progr. Theor. Phys. 96 (1996) 853.
\bi{gomm}
   H. Gomm, P. Jain, R. Johnson and J. Schechter, Phys. Rev. D 33 (1986) 3476;
   A. Migdal and M. Shifman,  Phys. Lett. B 114 (1982) 445.
\bi{meissmed}
   Ulf-G.Meissner, Nucl. Phys. A 503 (1989) 801.
\bi{anov}
   V.A. Andrianov and V.Yu. Novozhilov, Phys. Lett. B 202 (1988) 580;
   V. Nicolaev, O. Tkachev and V. Novozhilov, Nuov. Cim. A 107 (1994) 2674;
   V. Nicolaev and O. Tkachev, Phys. Elem. Part. and Nucl. 21 (1990) 1500.
\bi{our}
   M.M. Musakhanov and A. Rakhimov, Mod. Phys. Lett. A 10 (1995) 2297;
   A. Rakhimov, T. Okazaki,  M.M. Musakhanov and F.C.Khanna,
        Phys. Lett. B 378 (1996) 12.
\bi{kalfluid}
   G. Kalberman, Nucl. Phys. A 612  (1997) 359;
   G. Kalberman, "The Skyrmion in the Nucleus" (private communication).
\bi{bhaduri}
   Rajat K. Bhaduri, Models of nucleon from quarks to solitons,
        (Addison-Wiley publishing company INC 1988).
\bi{saitothomas}
   K. Saito, K. Tsushima and A. W. Thomas,
        Phys. Rev. C 55 (1997) 2637, and references therein.
\bi{verchuehara}
   H. Verchelde and H. Verbeke, Nucl. Phys. A 495  (1989) 523;
   A. Hayashi, S. Saito and M. Uehara, Phys. Rev. D 43 (1991) 1520.
\bi{workman}
   S. Nam, R.L. Workman,  Phys. Rev. D 41 (1990) 2323.
\bi{rho}
   G.E. Brown and M. Rho,  Phys. Rep.  269 (1996) 333.
\bi{birserev}
   M. Birse, J. Phys. G: Nucl. Part. Phys. 20 (1994) 1537.
\bi{ko}
   C.M. Ko, V. Koch and G. Li, Ann. Rev. Nucl. Part. Sci. 47 (1997) 505.
\bi{cohencg}
   T.D. Cohen, Phys. Rev. C 45 (1992) 1881.
\bi{kunihiro}
   T. Kunihiro, Progr. Theor. Phys. Suppl. 120 (1995) 75.
\bi{furnstahl}
   R.J. Furnstahl, X. Jin and D.B. Leinweber, Phys. Lett. B 387
   (1996) 253.
\bi{khannatowner}
   I.S. Towner and F.C. Khanna, Phys. Rev. Lett. 42  (1979) 51.
\bi{xzh}
   X. Zhu, S. Wong, F.C. Khanna, Y. Takahashi and T. Toyoda,
        Phys. Rev. C 36 (1987) 1968.
\bi{br}
   G.E. Brown and M. Rho, Phys. Rev. Lett. 66 (1991) 2720.
(See also ref. \ci{furnrho})
\bi{rappmach}
   R. Rapp, R. Machleidt, J.W. Durso and G.E. Brown,
        "Nuclear saturation with in - medium meson excange interactions",
       nucl-th/9706006.
\bi{hftheorem}
   T.D. Cohen, R.J. Furnstahl and D.K.Griegel,
        Phys. Rev. Lett. 67 (1991) 961.
\bi{reua}
   E. Reua, Rev. Mod. Phys. 46 (1974) 545.
\bi{akira}
   H. Ichie, A. Hatashigaki, A. Suzuki, M. Kumira,
        "Effective masses and sizes of $N(939)$, $\Delta (1232)$ and
        $N(1440)$ in nuclear medium", nucl-th/9308017.
\bi{vento}
   N.C. Mukhopadhyay and V. Vento,
        "On the Delta - Nucleon and Rho - Pi Splitting:
        A QCD - inspired Look in Free Hadrons versus Nuclei",
        nucl-th/9712073.
\bi{hatsudakunihiro}
   T.Hatsuda and T. Kunihiro, Phys. Lett. B 185 (1987) 304.
\bi{banerjeetjon}
   M.K. Banerjee and J.A. Tjon, Phys. Rev. C 56 (1997) 497.
\bi{birse96}
   M. Birse, Phys. Rev. C 53 (1996) R2048.
\bi{quarkmatter}
        Quark Matter'95, [Nucl. Phys. A 590 (1995)].
\bi{furnrho} C. Song, G.E. Brown, D. Min and M. Rho, Rhys. Rev.
C56 (1997) 2244;\\
R.J. Furnstahl, H. Tang and B. Serot Phys. Rev. C52 (1995) 1368
\bi{meisrep} U. - G. Meissner Phys. Rep. 161 (1988) 213.
\eb

\newpage
\centerline {\bf FIGURE CAPTIONS}
\begin{description}
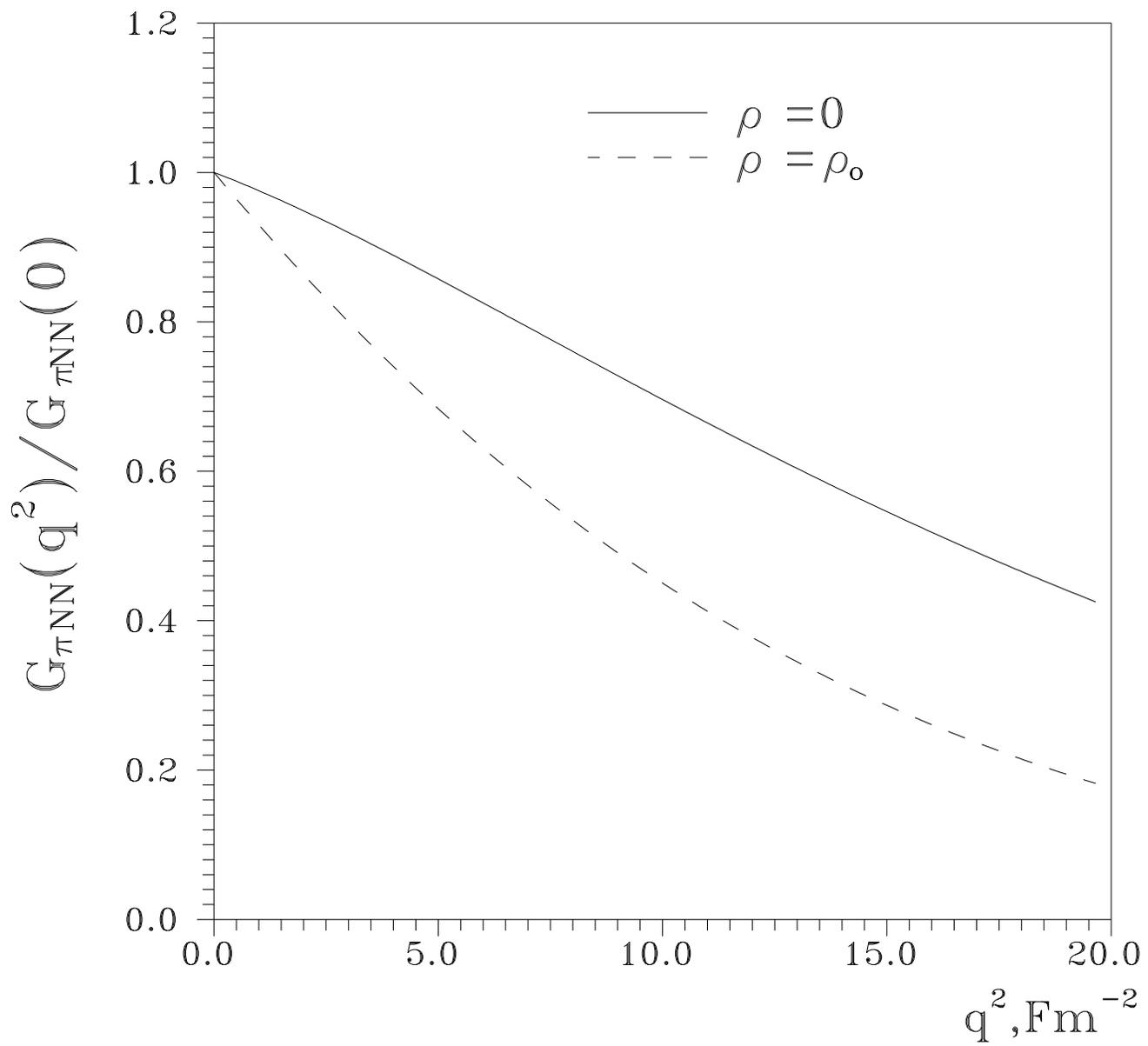
\item [Fig. 1.]
The $\pi NN$ form factor.The solid and dashed curves give results for
the free space ($\rho=0$) and the nuclear matter ($\rho=\rho_0)$
respectively.

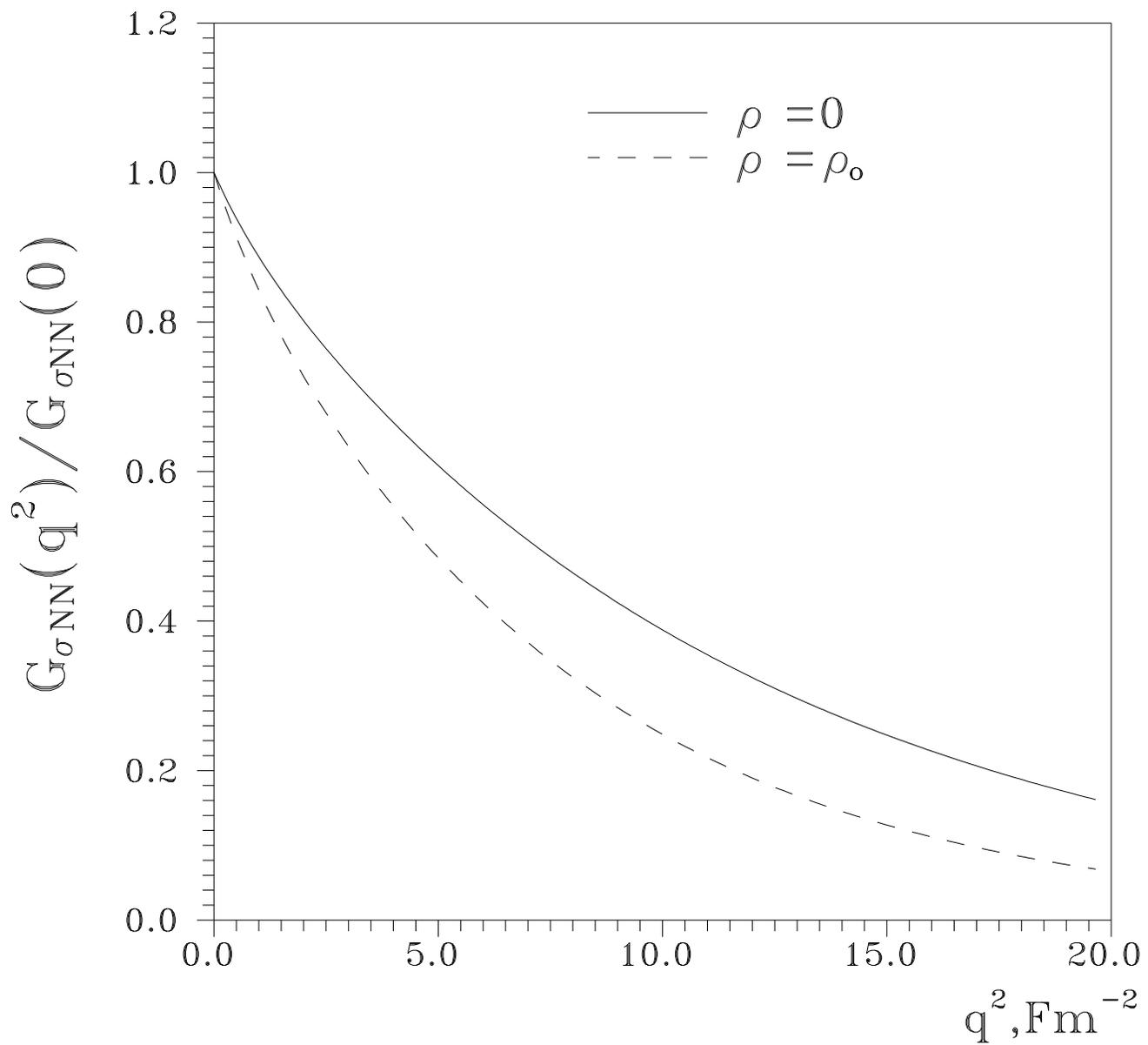
\item [Fig. 2.]
The $\sigma NN$ form factor.
Solid and dashed curves are for $\rho = 0$
and $\rho = \rho_0$ respectively.

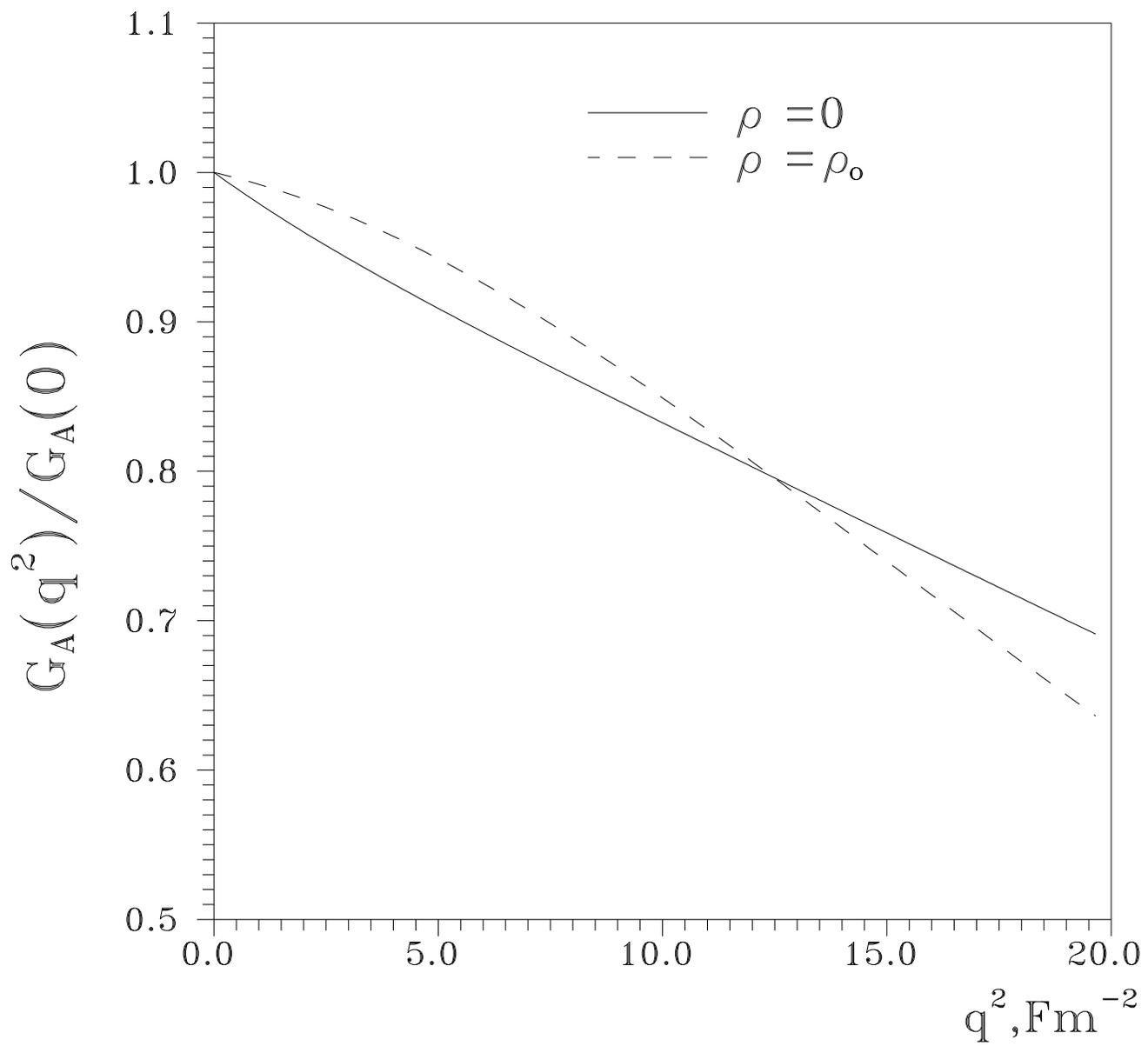
\item [Fig. 3.]
The axial form factor.
Solid and dashed curves are for $\rho = 0$
and $\rho = \rho_0$ respectively.

\item [Fig. 4.]
The profile functions $\Theta(r) $ (Fig. 4a.) and
$\sigma (r)$ (Fig. 4b.) for the $\rho=0$ (solid curve) and
$\rho=\rhozero$ (dashed curve). They are the solutions
 of equations of motion  \re{Eq}  in the sector with $B=1$.
\end{description}

\newpage

\begin{table}[tbp]
\caption{Density dependence of hadrons properties.
All quantities are in $MeV$.}
\bc
\begin{tabular}{cccccccc}
$\rho/\rho_0$&$ m_{\pi}^*$&$m_{\sigma}^*$&$(C_g^*)^{1/4}$&$M_N^*$&
$\Gamma_{\sigma\rightarrow\pi\pi}^*$&$\delta M_{N\Delta}^a$&
$\delta M_{N\Delta}^b$\\
\hline
0.0&139.00&550.1&260.70&1413&251.2&283.7&283.7\\
0.5&144.90&513.8&251.06&1271&88.7 &200.1&238.1\\
1.0&149.06&493.8&246.12&1157&34.6 &161.9&205.4
\end{tabular}
\ec
\end{table}

\begin{table} [tbp]
\caption{Coupling constants and cut-off parameters at finite density.
(All values are normalized to their free space ones.)}
\bc
\begin{tabular}{cccccccc}
$\rho/\rho_0$&
$g_{A}^*/g_{A}$&
$g_{\pi NN}^*/g_{\pi NN}$&
$g_{\sigma NN}^*/g_{\sigma NN}$&
$f_{\pi}^*/f_{\pi}$&
$\Lambda_{\pi}^*/\Lambda_{\pi}$&
$\Lambda_{\sigma}^*/\Lambda_{\sigma}$&
$r_a^*/r_a$\\
\hline
0.5&0.96&0.91&0.88&0.98&0.70&0.90&0.93\\
1.0&0.92&0.80&0.78&0.94&0.56&0.84&0.55
\end{tabular}
\ec
\end{table}

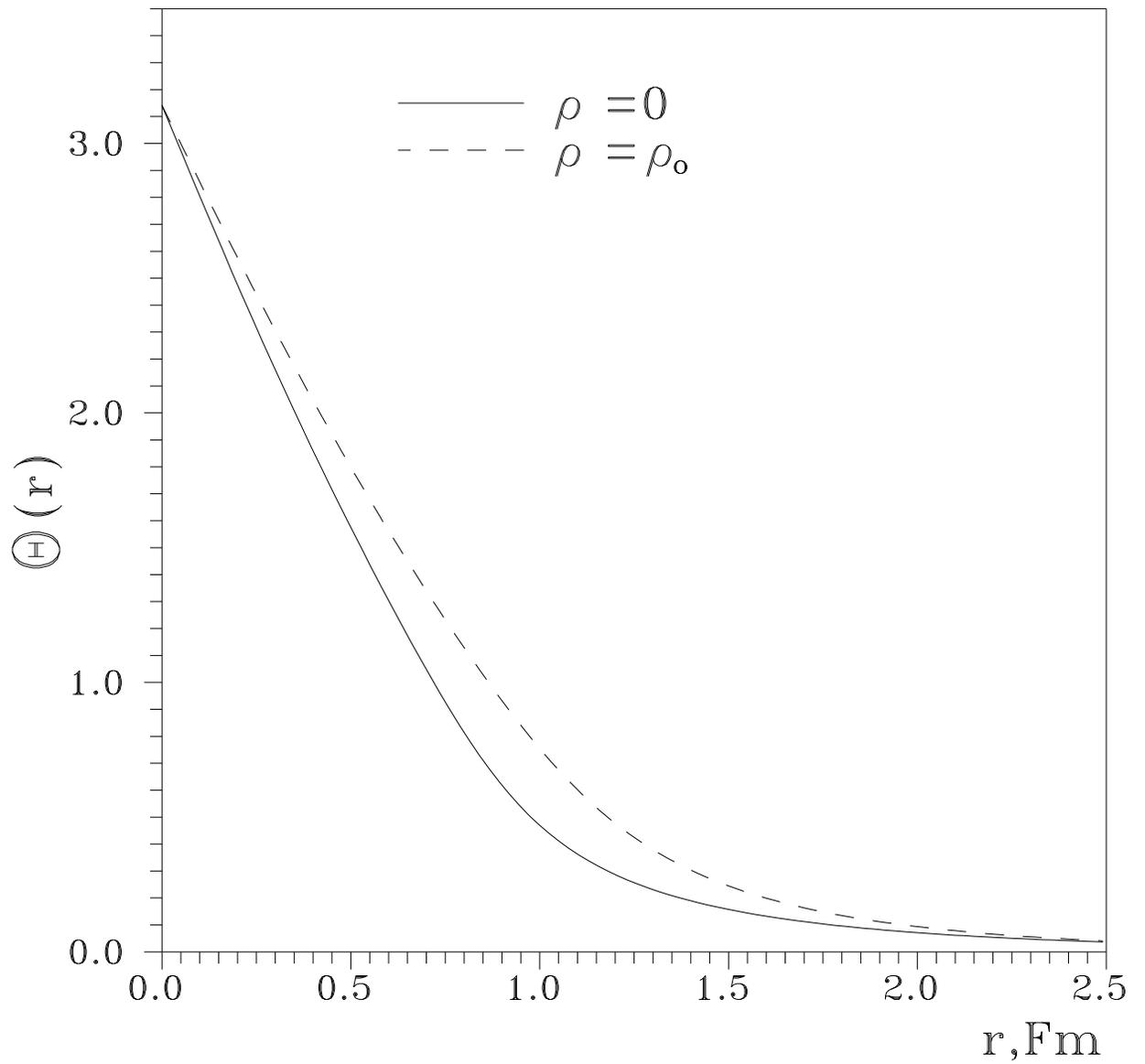
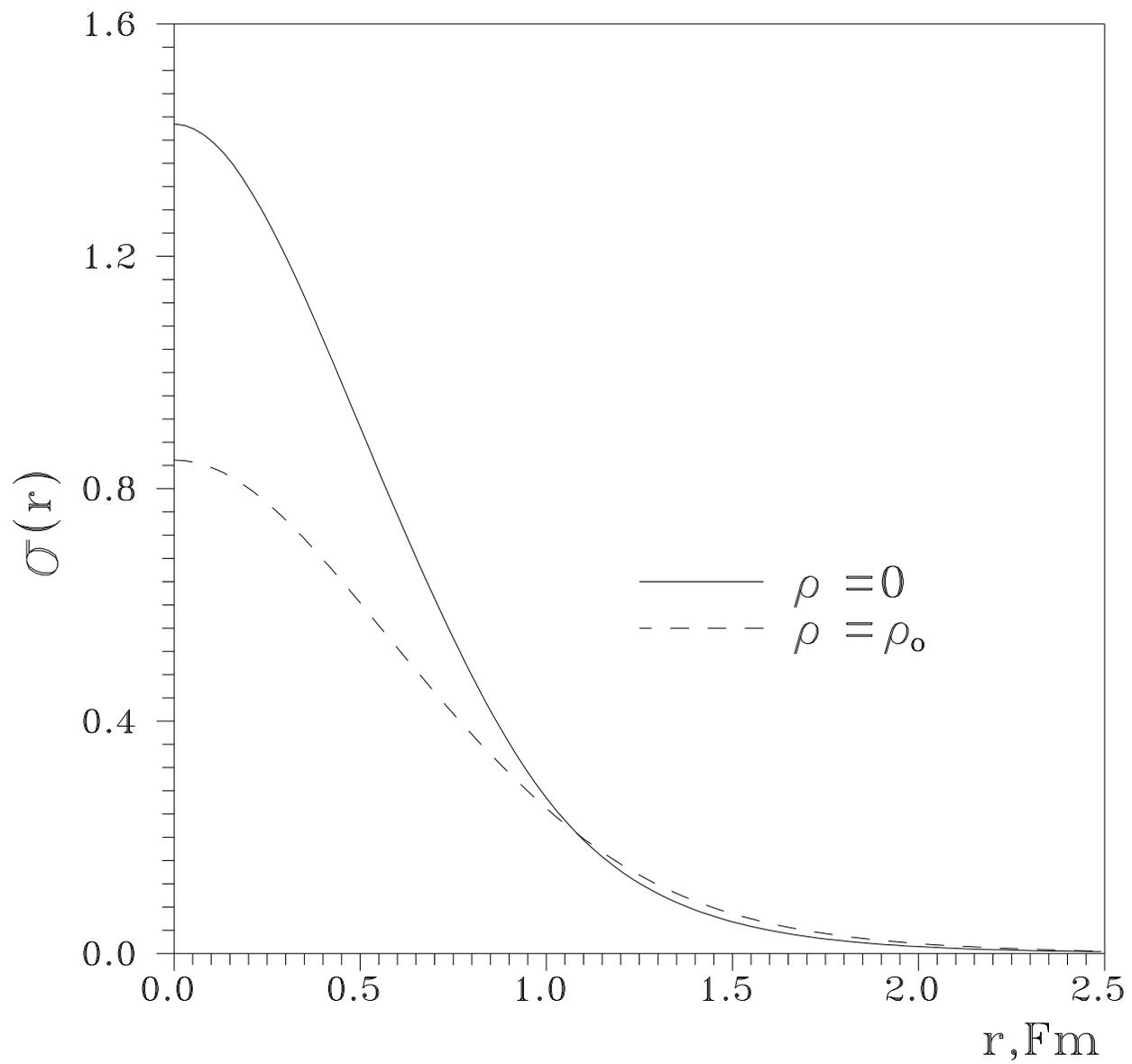
\end{document}